\documentclass{siamltex1213}
\usepackage{amsmath,amssymb,bm,graphicx}
\usepackage{float}
\restylefloat{table}
\usepackage{siunitx}
\usepackage{cite}
\usepackage[boxed]{algorithm2e}

\newcolumntype{?}{!{\vrule width 1pt}}
  
\title{Varying the resolution of the Rouse model on temporal 
and spatial scales: application to multiscale modelling of DNA dynamics}
\author{Edward Rolls\footnotemark[1]
\and Yuichi Togashi\footnotemark[2]
\and Radek Erban\footnotemark[1]}

\begin{document}
\maketitle
\slugger{mms}{xxxx}{xx}{x}{x--x}

\renewcommand{\thefootnote}{\fnsymbol{footnote}}

\footnotetext[1]{Mathematical Institute, University of Oxford, 
Radcliffe Observatory Quarter, Woodstock Road, Oxford OX2 6GG, United Kingdom; 
e-mails: edward.rolls@pmb.ox.ac.uk, erban@maths.ox.ac.uk. Edward Rolls 
and Radek Erban would like to thank the Isaac Newton Institute for 
Mathematical Sciences, Cambridge, for support and hospitality during 
the programme ``Stochastic Dynamical Systems in Biology: 
Numerical Methods and Applications" where work on this paper was partially
undertaken. This work was supported by EPSRC grants no EP/K032208/1 
and EP/G03706X/1. This work was supported by a grant from the Simons 
Foundation. Radek Erban would also like to thank the Royal Society for 
a University Research Fellowship.
}
\footnotetext[2]{Research Center for the Mathematics on Chromatin 
Live Dynamics, Hiroshima University,
1-3-1 Kagamiyama, Higashi-Hiroshima, Hiroshima 739-8526, Japan;
e-mail: togashi@hiroshima-u.ac.jp.
This work was partially supported by KAKENHI 23115007 
and 16H01408, MEXT, Japan,
and Platform for Dynamic Approaches to Living System, AMED, Japan.}

\renewcommand{\thefootnote}{\arabic{footnote}}

\begin{abstract}
A multi-resolution bead-spring model for polymer dynamics is developed
as a generalization of the Rouse model. A polymer chain is described
using beads of variable sizes connected by springs with variable 
spring constants. A numerical scheme which can use different 
timesteps to advance the positions of different beads is presented 
and analyzed. The position of a particular bead is only updated at 
integer multiples of the timesteps associated with its connecting 
springs. This approach extends the Rouse model to a multiscale 
model on both spatial and temporal scales, allowing simulations of
localized regions of a polymer chain with high spatial and temporal
resolution, while using a coarser modelling approach to describe 
the rest of the polymer chain. A method for changing the model 
resolution on-the-fly is developed using the Metropolis-Hastings 
algorithm. It is shown that this approach maintains key 
statistics of the end-to-end distance and diffusion of the polymer
filament and makes computational savings when applied to a model 
for the binding of a protein to the DNA filament.
\end{abstract}


\begin{AM}60H10, 60J70, 82C31, 82D60, 92B99\end{AM}

\pagestyle{myheadings}
\thispagestyle{plain}
\markboth{Edward J. Rolls, Yuichi Togashi, Radek Erban}{Multiscale 
Modelling of DNA Dynamics}

\section{Introduction}
Over the past 70 years, there have been multiple attempts to 
dynamically model the movement of polymer chains with Brownian
dynamics~\cite{Rotne1969tensor,rouse1953model,Zimm1959Model,kremer1990dynamics},
which have more recently been used as a model for 
DNA filament dynamics~\cite{Andrews2014methods}. One of the first 
and simplest descriptions was given as the Rouse 
model~\cite{rouse1953model}, which is a bead-spring
model~\cite{Andrews2014methods}, where the continuous filament 
is modelled at a mesoscopic scale with beads connected by springs. 
The only forces exerted on beads are spring forces from adjacent 
springs, as well as Gaussian noise. Hydrodynamic forces between 
beads and excluded volume effects are neglected in the model 
in favour of simplicity and computational speed, but the model 
manages to agree with several properties of polymer chains from
experiments~\cite{kremer1990dynamics,richter1981dynamics}. Other 
models exist, for example the Zimm model introduces hydrodynamic 
forces~\cite{Zimm1959Model} between beads, or bending potentials 
can be introduced to form a wormlike chain and give a notion of 
persistence length~\cite{allison1986brownian}, see, for example, 
review article~\cite{Andrews2014methods} or 
books~\cite{Doi1986theory,de1979scaling} on this subject.

Most of the aforementioned models consider the filament on only 
a single scale. In some applications, a modeller is 
interested in a relatively small region of a complex system.
Then it is often possible to use a hybrid model which is more 
accurate in the region of interest, and couple this with a model 
which is more computationally efficient in the rest of the 
simulated domain~\cite{flegg2012two,erban2014molecular,couplingallatom}.
An application area for hybrid models of polymer chains is 
binding of a protein to the DNA filament, which 
we study in this paper. The model which we have created uses 
Rouse dynamics for a chain of DNA, along with a freely diffusing 
particle to represent a binding protein. As the protein approaches 
the DNA, we increase the resolution in the 
nearby DNA filament to increase accuracy of our simulations, 
whilst keeping them computationally efficient.  

In this paper we use the Rouse model for analysis due to its mathematical
tractability and small computational load. Such a model is applicable
to modelling DNA dynamics when we consider relatively low resolutions,
when hydrodynamic forces are negligible and persistence length is 
significantly shorter than the Kuhn length between 
each bead~\cite{Andrews2014methods}. 
The situation becomes more complicated when we consider DNA modelling
at higher spatial resolutions. 

Inside the cell nucleus, genetic information is stored within strands 
of long and thin DNA fibres, which are separated into chromosomes. 
These DNA fibres are folded into structures related to their function. 
Different genes can be enhanced or inhibited depending upon this
structure~\cite{felsenfeld2003controlling}. Folding also minimises 
space taken up in the cell by DNA~\cite{kornberg1974chromatin}, and 
can be unfolded when required by the cell for different stages in 
the cell cycle or to alter gene expression. The folding of DNA occurs 
on multiple scales. On a microscopic scale, DNA is wrapped around 
histone proteins to form the nucleosome structure~\cite{wolffe2000chromatin}.
This in turn gets folded into a chromatin fibre which gets packaged 
into progressively higher order structures until we reach the level 
of the entire chromosome~\cite{felsenfeld2003controlling}. The finer 
points of how the nucleosome packing occurs on the chromatin fibre and 
how these are then packaged into higher-order structures is still 
a subject of much debate, with long-held views regarding mesoscopic 
helical fibres becoming less fashionable in favour of more 
irregular structures in vivo~\cite{maeshima2016}. 

In the most compact form of chromatin, many areas of DNA are 
not reachable for vital reactions such as
transcription~\cite{felsenfeld2003controlling}. One potential 
explanation to how this is overcome by the cell is to position 
target DNA segments at the surface of condensed domains when 
it is needed~\cite{maeshima2015physical,cremer2015review}, 
so that transcription 
factors can find expressed genes without having to fit into these 
tightly-packed structures. This complexity is not captured by the 
multiscale model of protein binding presented in this
paper. However, if one uses the developed refinement of the Rouse model
together with a more detailed modelling approach in a small
region of DNA next to the binding protein, then such
a hybrid model can be used to study the effects of 
microscopic details on processes over system-level spatial 
and temporal scales. When taking this multiscale 
approach, it is necessary to understand the error from 
including the less accurate model in the hybrid model 
and how the accuracy of the method depends on its parameters.
These are the main questions studied in this paper.

The rest of the paper is organized as follows. In 
Section~\ref{secmrbs}, we introduce a multi-resolution 
bead-spring model which generalizes the Rouse model.
We also introduce a discretized version of this model
which enables the use of different timesteps in different
spatial regions. In Section~\ref{section3}, we analyze
the main properties of the multi-resolution bead-spring
model. We prove two main lemmas giving formulas
for the diffusion constant and the end-to-end distance.
We also study the appropriate choice of timesteps for
numerical simulations of the model and support our analysis
by the results of illustrative computer simulations.
Our main application area is studied in Section~\ref{section4}
where we present and analyze a DNA binding model. We  
develop a method to increase the resolution in existing segments 
on-the-fly using the Metropolis-Hastings algorithm.
In Section~\ref{secdiscussion}, we conclude our paper by 
discussing possible extensions of the presented multiscale 
approach (by including more detailed models of DNA
dynamics) and other multiscale methods developed in 
the literature.

\section{Multi-resolution bead-spring model}

\label{secmrbs}

We generalize the classical Rouse bead-spring polymer
model~\cite{rouse1953model} to include 
beads of variable sizes and springs with variable spring constants. In 
Definition~\ref{defmrbs}, we formulate the evolution equation for
this model as a system of stochastic differential equations (SDEs). 
We will also introduce a discretized version of this model in 
Algorithm~\ref{algoneiter}, which will be useful in 
Sections~\ref{section3}~and~\ref{section4}
where we use the multi-resolution bead-spring model to  
develop and analyze multiscale models for DNA dynamics.

\smallskip

\begin{definition}
\label{defmrbs}
Let $N>1$ be a positive integer. A multi-resolution bead-spring 
polymer of size $N$ consists of a chain of $N$ beads of radius
$\sigma_n$, for $n = 1,2,\dots,N$, connected by $N-1$ springs which 
are characterized by their spring constants $k_n$, for $n=1,2,\dots,N-1$. 
The positions $\mathbf{r}_n \equiv \mathbf{r}_n(t) 
= [r_{n,1}(t),r_{n,2}(t),r_{n,3}(t)]$
of beads evolve according to the system of SDEs (for $n = 1,2,\dots,N$)
\begin{equation}
\zeta_n \, \mbox{{\rm d}}\mathbf{r}_n 
=
\Big( k_{n-1} \mathbf{r}_{n-1}(t) - (k_{n-1}+k_n) \mathbf{r}_n(t) 
+ k_{n} \mathbf{r}_{n+1}(t) \Big) \, \mbox{{\rm d}}t
+ 
\sqrt{2 k_B T \zeta_n}
\; \mbox{{\rm d}}{\mathbf W}_n,
\label{SDEdef}
\end{equation}
where $\zeta_n = 6 \pi \eta \sigma_n$ is the frictional drag 
coefficient of the $n$-th bead given by Stokes' Theorem, $\eta$ 
is the solvent viscosity, $\mbox{{\rm d}}{\mathbf W}_n \equiv 
[\mbox{{\rm d}}W_{n,1},\mbox{{\rm d}}W_{n,2}, \mbox{{\rm d}}W_{n,3}]$ 
is a Wiener process, $T$ is absolute temperature, $k_B$ 
is Boltzmann's constant and we assume that each spring constant
$k_n$ can be equivalently expressed in terms 
of the corresponding Kuhn length $b_n$ by
$$
k_n = \frac{3k_B T}{b_{n}^2}, 
\qquad \mbox{for} \quad 
n=1,2,\dots,N-1.
$$
We assume that the behaviour of boundary beads (for $n=1$ and $n=N$) is 
also given by equation $(\ref{SDEdef})$ simplified by postulating 
$\mathbf{r}_0(t) = \mathbf{r}_1(t)$ and
$\mathbf{r}_{N+1}(t) = \mathbf{r}_N(t).$
\end{definition}

\begin{figure}
\centering
\includegraphics[width=\linewidth]{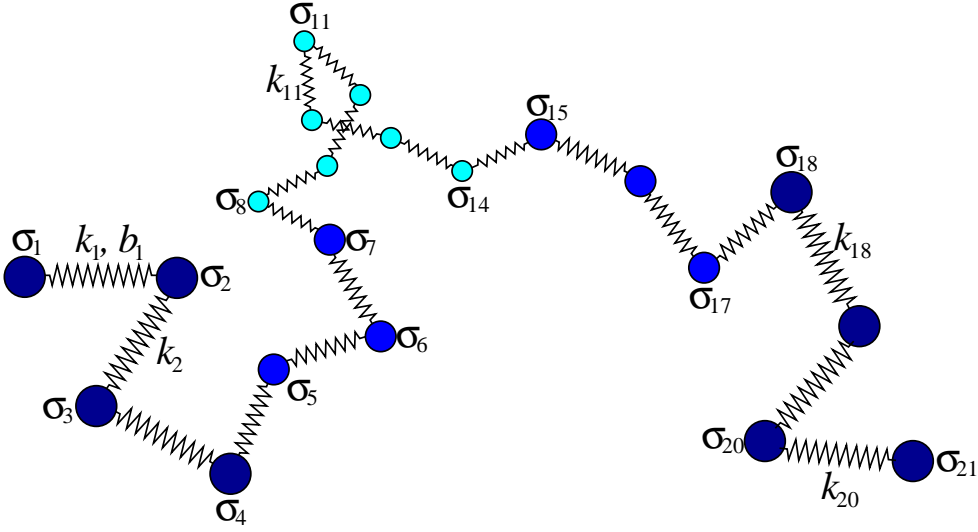}
\caption{{\it 
A schematic of the multi-resolution bead-spring model for $N=21.$}
\label{figmrbeadspring}}
\end{figure}

\medskip

\noindent
In Figure~\ref{figmrbeadspring}, we schematically illustrate a
multi-resolution bead-spring polymer for $N=21$. The region
between the $8$-th and the $14$-th bead is described with 
the highest resolution by considering smaller beads and springs 
with larger spring constants
(or equivalently with smaller Kuhn lengths). The scalings of
different parameters in Definition~\ref{defmrbs} are chosen so
that we recover the classical Rouse model~\cite{rouse1953model}
if we assume $\sigma_1 = \sigma_2 = \dots = \sigma_N = \sigma$
and $b_1 = b_2 = \dots = b_{N-1} = b$. Then equation (\ref{SDEdef})
simplifies to
\begin{equation}
\zeta \, \mbox{{\rm d}}\mathbf{r}_n 
=
k \, \Big(\mathbf{r}_{n-1}(t) - 2 \mathbf{r}_n(t) 
+ \mathbf{r}_{n+1}(t) \Big) \, \mbox{{\rm d}}t
+ 
\sqrt{2 k_B T \zeta}
\; \mbox{{\rm d}}{\mathbf W}_n,
\label{SDERouse}
\end{equation}
where $\zeta = 6 \pi \eta \sigma$, $k  = 3k_B T / b^2$ and
we again define $\mathbf{r}_0(t) = \mathbf{r}_1(t)$ and
$\mathbf{r}_{N+1}(t) = \mathbf{r}_N(t)$ in equations for
boundary beads. In the polymer physics literature~\cite{Doi1986theory}, 
the Rouse model (\ref{SDERouse}) is equivalently written as 
\begin{equation}
\label{SDERouse2}
\zeta \frac{d \mathbf{r}_n}{dt}(t) 
= k \, \Big(\mathbf{r}_{n-1}(t) - 2 \mathbf{r}_n(t) 
+ \mathbf{r}_{n+1}(t) \Big) 
+ \mathbf{f}_n(t),
\end{equation}
where random thermal noises $\mathbf{f}_n(t)$ 
exerted on the beads from Brownian motion are characterized by the 
moments~\cite{Doi1986theory}
\begin{equation}
\langle \mathbf{f}_n (t) \rangle = \mathbf{0}, \qquad
\langle f_{n,i} (t) f_{m,j} (t') \rangle 
= 2 \zeta k_B T \delta_{nm}\delta_{ij} \delta(t-t'),
\label{moments}
\end{equation}
where $n,m=1,2,\ldots,N$ and $i,j \in \{1,2,3\}$. 
For the remainder of this paper, we will use the SDE notation
as given in~(\ref{SDERouse}), 
because we will often study numerical schemes for simulating polymer 
dynamics models. The simplest discretization of (\ref{SDERouse}) is 
given by the Euler-Maruyama method~\cite{kloeden1992}, which 
uses the finite timestep $\Delta t$ and calculates the position 
vector $\mathbf{r}_n^t$ of the $n$-th bead, $n=1,2,\ldots,N$, 
at discretised time $t$ by
\begin{equation}
\mathbf{r}_n^{t} = 
\mathbf{r}_n^{t-\Delta t} 
+ \frac{k}{\zeta}
\left(
\mathbf{r}_{n-1}^{t-\Delta t}
- 
2 \, \mathbf{r}_{n}^{t-\Delta t} 
+
\mathbf{r}_{n+1}^{t-\Delta t}
\right) \, \Delta t 
+ \sqrt{\frac{2 k_B T}{\zeta} \Delta t} \ \bm{\xi}_n
\label{EMdiscR}
\end{equation}
for $\bm{\xi}_n=(\xi_{n,1},\xi_{n,2},\xi_{n,3})$, 
where $\xi_{n,i}$ is normally distributed random variable with
zero mean and unit variance (i.e. $\xi_{n,i} \sim \mathcal{N}(0,1)$) 
for $i = 1,2,3$. In order to discretize the multi-resolution 
bead-spring model, we allow for variable timesteps.

\smallskip

\begin{definition}
\label{defvartimestep}
Let $\Delta t > 0$ and let $j_n$, $n=1,2,\dots,N-1,$ be positive integers 
such that $j_{n-1} \mid j_{n}$ or $j_{n} \mid j_{n-1}$ for $n=2,3,\dots,N-1$.
Let us assume that at least one of the values of $j_n$ is equal to 1.
We define $\Delta t_n = j_n \Delta t$ for $n=1,2,\dots,N-1$ and 
we call $\Delta t_n$ a timestep associated with the $n$-th spring.
\end{definition}

\medskip

\noindent
Definition~\ref{defvartimestep} specifies that all timesteps must 
be integer multiples of the smallest timestep $\Delta t$. 
The timesteps associated with two adjacent springs are also
multiples of each other. The time evolution of the multi-resolution
bead-spring model is computed at integer multiples of $\Delta t$.
One iteration of the algorithm is shown in Algorithm~\ref{algoneiter}.
The position of the $n$-th bead is updated at integer multiples of 
$\min\{j_{n-1},j_n\} \Delta t = \min\{\Delta t_{n-1}, \Delta t_n \}$
by calculating the random displacement due to Brownian motion, 
with displacement caused by springs attached to the bead also updated 
at integer multiples of the timesteps associated with each spring, 
i.e. $\Delta t_{n-1}$ or $\Delta t_n.$ Considering the situation that 
all beads, springs and timesteps are the same, then one can easily 
deduce the following result.

\begin{algorithm}[t]
Update positions of internal beads which are connected to two springs: \\
 \For{$n=2,3,\dots,N-1$}{
  \eIf{$\min\{j_{n-1},j_n\} \mid i$} {
   Put $\mathbf{r}_n^{i \Delta t} := \mathbf{r}_n^{(i-1) \Delta t}
   + \sqrt{\displaystyle \frac{2 k_B T}{\zeta_n} 
   \min\{j_{n-1},j_n\} \Delta t } \, \bm{\xi}_n^i$
   where $\bm{\xi}_n^i$ is a vector \\ of
   random numbers sampled from 
   $\bm{\xi}_n^i \sim [\mathcal{N}(0,1),\mathcal{N}(0,1),\mathcal{N}(0,1)]$. \\
   \If{$j_{n-1} \mid i$}{
   Put $\mathbf{r}_n^{i \Delta t} := \mathbf{r}_n^{i \Delta t} 
    + \displaystyle \frac{k_{n-1}}{\zeta_n}
   \left( \mathbf{r}_{n-1}^{(i-j_{n-1}) \Delta t} 
   - \mathbf{r}_{n}^{(i-j_{n-1}) \Delta t} \right) \, \Delta t_{n-1}$.}
   \If{$j_{n} \mid i$}{
   Put $\mathbf{r}_n^{i \Delta t} := \mathbf{r}_n^{i \Delta t} 
    + \displaystyle \frac{k_n}{\zeta_n}
   \left( \mathbf{r}_{n+1}^{(i-j_{n}) \Delta t} 
   - \mathbf{r}_{n}^{(i-j_{n}) \Delta t} \right) \, \Delta t_{n}$.} }
   {Put $\mathbf{r}_n^{i \Delta t} := \mathbf{r}_n^{(i-1) \Delta t}$.} 
   }
   \smallskip
   Update of the first bead: \\
  \eIf{$j_1 \mid i$} {
   Put $\mathbf{r}_1^{i \Delta t} := \mathbf{r}_1^{(i-1) \Delta t}
   + \displaystyle \frac{k_1}{\zeta_1}
   \left( \mathbf{r}_{2}^{(i-j_1) \Delta t} 
   - \mathbf{r}_{1}^{(i-j_1) \Delta t} \right) \, \Delta t_1
   + 
   \mbox{$\sqrt{\displaystyle \frac{2 k_B T}{\zeta_1} \Delta t_1 }$} 
   \, \bm{\xi}_1^i$ \\
   where $\bm{\xi}_1^i \sim
   [\mathcal{N}(0,1),\mathcal{N}(0,1),\mathcal{N}(0,1)]$. }
   {Put $\mathbf{r}_1^{i \Delta t} := \mathbf{r}_1^{(i-1) \Delta t}$.}
   \smallskip
   Update of the last bead: \\ 
  \eIf{$j_{N-1} \mid i$} {
   Put $\mathbf{r}_N^{i \Delta t} := \mathbf{r}_N^{(i-1) \Delta t}
   + \displaystyle \frac{k_{N-1}}{\zeta_N}
   \left( \mathbf{r}_{N-1}^{(i-j_{N-1}) \Delta t}
   - \mathbf{r}_{N}^{(i-j_{N-1}) \Delta t} \right) \, \Delta t_{N-1}$ \\
   \rule{0pt}{0pt} \hskip 1.5 cm $+ \, 
   \sqrt{\displaystyle \frac{2 k_B T}{\zeta_N} \Delta t_{N-1} } 
   \, \bm{\xi}_N^i$
   where $\bm{\xi}_N^i \sim
   [\mathcal{N}(0,1),\mathcal{N}(0,1),\mathcal{N}(0,1)]$. }
   {Put $\mathbf{r}_N^{i \Delta t} := \mathbf{r}_N^{(i-1) \Delta t}$.} 
\caption{{\it One\hbox{\strut}
iteration of the numerical algorithm for simulating 
the multi-resolution bead-spring model introduced in 
Definition~$\ref{defmrbs}$. We update positions of beads 
at time $t = i \Delta t$ where $i$ is a positive integer and
each spring is simulated using its associated timestep given 
in Definition~$\ref{defvartimestep}$.}
\label{algoneiter}}
\end{algorithm}

\smallskip

\begin{lemma}
\label{lemconsnum}
Let $\sigma > 0$, $\zeta > 0$, $k > 0$ and $\Delta t > 0$ be positive 
constants and $N>1$ be an integer. Consider a multi-resolution 
bead-spring polymer of size $N$ with $\sigma_n = \sigma$, 
$\zeta_n = \zeta$, for $n = 1,2,\dots,N$, and $k_n = k$, for $n=1,2,\dots,N-1$. 
Let the timesteps associated with each spring be equal to
$\Delta t$, i.e. $j_1=j_2=\dots=j_n=1$ in Definition~$\ref{defvartimestep}.$ 
Then Algorithm $\ref{algoneiter}$ is equivalent to the Euler-Maruyama 
discretization of the Rouse model given as equation $(\ref{EMdiscR})$.
\end{lemma}

\medskip

\noindent
Lemma~\ref{lemconsnum} shows that the multi-resolution bead-spring model
is a generalization of the Rouse model. In the next section, we will
study properties of this model which will help us to select the
appropriate parameter values for this model and use it in multiscale
simulations of DNA dynamics.

\section{Macroscopic properties and parameterizations of 
multi-resolution bead-spring models}

\label{section3}

We have formulated a multiscale Rouse model which varies the Kuhn 
lengths throughout the filament, but we would like to keep properties 
of the overall filament constant regardless of the resolution regime 
being considered for the filament. We consider a global statistic 
for the system to be \textit{consistent} if the expected value of 
the statistic is invariant to the resolution regime being considered 
for the filament. We consider the
\textit{self diffusion constant} and 
\textit{root mean squared (rms) end-to-end distance}
as two statistics we wish to be consistent in our system, 
which can be ensured by varying the bead radius and the 
number of beads respectively. The precise way to vary these properties will 
be explored in this section.

\subsection{Self diffusion constant}
\label{secdiffusion}
The \textit{self diffusion constant} is defined as
\begin{equation}
\label{DGdef}
D_G 
= \lim_{t \rightarrow \infty} \frac{1}{6t} 
\langle (\mathbf{r}_G^t - \mathbf{r}_G^0)^2 \rangle 
= \lim_{t \rightarrow \infty} \frac{1}{6t} \sum_{i=1}^3 
\langle (r_{G,i}^t - {r}_{G,i}^0)^2 \rangle,
\end{equation}
where $\mathbf{r}_G^t$ is the \textit{centre of mass} of the polymer 
chain at time $t$, which is defined by
\begin{equation}
\label{eq:COM}
\mathbf{r}_G^t = \frac{1}{\Omega} \sum_{n=1}^N \sigma_{n} \mathbf{r}_n^t,
\qquad
\mbox{where}
\qquad
\Omega = \sum_{n=1}^N \sigma_{n}.
\end{equation}
Definition (\ref{eq:COM}) is an extension to the definition given 
by Doi and Edwards~\cite{Doi1986theory} for the centre of mass of a continuous 
chain on only one scale. If all beads have the same radius $\sigma$
(i.e. if $\sigma_n = \sigma$ for $n=1,2,\dots,N$), then equation
(\ref{eq:COM}) simplifies to the centre of mass definition for
the classical Rouse model. In this case, the self diffusion
constant is given by~\cite{Doi1986theory}
\begin{equation}
\label{eq:rouseDiff}
D_G = \frac{k_B T}{6 \pi \eta \sigma N},
\end{equation}
where $N$ is the number of beads. This result explains the, on the face of it,
counterintuitive scaling of equation (\ref{eq:COM}) with $\sigma_n$.
If we suppose that each bead had the same density, then the mass
of each bead would be proportional to its volume, i.e. to $\sigma_n^3$.
However, in definition (\ref{eq:COM}), we have used weights $\sigma_n$
instead of $\sigma_n^3,$ because beads do not represent physical bead 
objects like nucleosomes, but representations of the filament around it, 
so the bead radius scales with the amount of surrounding filament, 
which is linear in bead radius in this formulation. 
If we consider DNA applications, we could imagine each bead as 
a tracker for individual base pairs at intervals of, say,
thousands of base pairs away from each other along the DNA filament. 
The filament in the model is then drawn between adjacent beads.

This linear scaling with $\sigma_n$ can also be confirmed using
equation~(\ref{eq:rouseDiff}) for the classical Rouse model. 
If we describe the same polymer using a more detailed model consisting of twice 
as many beads (i.e. if we change $N$ to $2N$), then we have to halve 
the bead radius (i.e. change $\sigma$ to $\sigma/2$) to get 
a polymer model with the same diffusion constant~(\ref{eq:rouseDiff}). 
In particular, the mass of a bead scales with $\sigma$ (and not 
with $\sigma^3$). In the next lemma, we extend result~(\ref{eq:rouseDiff}) 
to a general multi-resolution bead-spring model.

\smallskip

\begin{lemma}
\label{lemdg}
Let us consider a multi-resolution bead-spring polymer of size $N$
and a set of timesteps associated with each spring satisfying
the assumptions of Definitions~$\ref{defmrbs}$ and 
$\ref{defvartimestep}$. Then the self diffusion constant
of the polymer evolution described by Algorithm~$\ref{algoneiter}$
is given by
\begin{equation}
\label{eq:SDC}
 D_G = \frac{k_B T}{6 \,\! \pi \,\! \eta \, \Omega}.
\end{equation}
\end{lemma}

\smallskip

\begin{proof}
Algorithm~$\ref{algoneiter}$ describes one iteration of our numerical
scheme. Multiplying the steps corresponding to the $n$-th bead 
by $\sigma_n$ and summing over all beads, we obtain how 
$\Omega \, \mathbf{r}_G^{i \Delta t}$ changes during one timestep
$\Delta t$. Since $\zeta_n = 6 \pi \eta \sigma_n$, tension terms
cancel after summation and the evolution rule for 
$\Omega \, \mathbf{r}_G^{i \Delta t}$ simplifies to
\begin{eqnarray}
\Omega \, \mathbf{r}_G^{i \Delta t}
&=&
\Omega \, \mathbf{r}_G^{(i-1) \Delta t}
+
Q(j_1,i)
\sqrt{\displaystyle \frac{k_B T \sigma_1}{3 \pi \eta} \Delta t_1 } 
\, \bm{\xi}_1^i
+
Q(j_{N-1},i)
\sqrt{\displaystyle \frac{k_B T \sigma_N}{3 \pi \eta} \Delta t_{N-1} } 
\, \bm{\xi}_N^i 
\nonumber
\\
&+& 
\sum_{n=2}^{N-1}
Q(\min\{j_{n-1},j_n\},i)
\sqrt{\displaystyle 
\frac{k_B T \sigma_n}{3 \pi \eta} 
\min\{j_{n-1},j_n\} \Delta t } \, \bm{\xi}_n^i,
\label{rgequation}
\end{eqnarray}
where 
$\bm{\xi}_n^i \sim [\mathcal{N}(0,1),\mathcal{N}(0,1),\mathcal{N}(0,1)]$ and 
function $Q(j,i)$ is defined for positive integers $j$ and $i$ by
$$
Q(j,i) 
= 
\left\{ \begin{matrix} 1, & \quad \mbox{if} \; j \mid i; \\ 
                                0, & \quad \mbox{if} \; j \nmid i.
		 \end{matrix} \right.
$$		 
Let us denote by $H$ the least common multiple of $\{j_1, j_2, \dots, j_n\}.$
Every bead is updated in Algorithm~$\ref{algoneiter}$ at integer multiples
of $H \Delta t$. We can eliminate function $Q$ from equation (\ref{rgequation})
if we consider the evolution of $\Omega \, \mathbf{r}_G^{t}$ when time $t$ is
evaluated at integer multiples of $H \Delta t$. We obtain the evolution rule
\begin{equation}
\Omega \, \mathbf{r}_G^{i H \Delta t}
=
\Omega \, \mathbf{r}_G^{(i-1) H \Delta t}
+
\sqrt{\displaystyle \frac{k_B T \Omega}{3 \pi \eta} H \Delta t } \, \bm{\xi},
\label{rgequation2}
\end{equation}
for $\bm{\xi} \sim [\mathcal{N}(0,1),\mathcal{N}(0,1),\mathcal{N}(0,1)]$,
where we used the fact that the sum of normally distributed random variables
is again normally distributed. Dividing equation (\ref{rgequation2}) by
$\Omega$, we obtain
$$
\mathbf{r}_G^{i H \Delta t}
=
\mathbf{r}_G^{0}
+
\sqrt{\displaystyle \frac{k_B T}{3 \pi \eta \, \Omega} i H \Delta t } 
\, \bm{\xi},
\qquad \mbox{for} \quad \bm{\xi} 
\sim [\mathcal{N}(0,1),\mathcal{N}(0,1),\mathcal{N}(0,1)].
$$
Using definition (\ref{DGdef}), we obtain (\ref{eq:SDC}).
\end{proof}

\medskip

\noindent
The formula (\ref{eq:SDC}) is a generalization of equation
(\ref{eq:rouseDiff}) obtained for the Rouse model. It is 
invariant to the resolutions provided that the mass of the 
filament $\Omega$ remains constant through selection of the 
number of beads and bead radius, therefore the self diffusion 
constant is consistent.

\subsection{End-to-end distance}
\label{sec:noballs}

We define the \textit{end-to-end vector} 
$\mathbf{R} = \mathbf{r}_N - \mathbf{r}_1$ from 
one end of the filament to the other~\cite{Doi1986theory}. 
An important statistic to consider related to this is 
the \textit{root mean squared (rms) end-to-end distance} of 
the filament $\mu = \langle \mathbf{R}^2 \rangle^{1/2}$. The 
expected value of the long-time limit of the rms end-to-end distance, 
denoted $\mu_{\infty}$, for the classical Rouse model is given 
by~\cite{Doi1986theory}
\begin{equation}
\label{eq:rmse2e}
\mu_{\infty} 
= \lim_{t \rightarrow \infty} \langle \mathbf{R}^2 \rangle^{1/2} 
= b \sqrt{N-1}.
\end{equation}
We generalize this result in the following lemma.

\smallskip

\begin{lemma}
\label{lemrms}
Let us consider a multi-resolution bead-spring polymer of size 
$N$ satisfying the assumptions of Definition~$\ref{defmrbs}$. 
Then 
\begin{equation}
\label{rmsbond}
\lim_{t \rightarrow \infty} 
\langle 
\left(\mathbf{r}_{n+1}(t) - \mathbf{r}_{n}(t) \right)^2 \rangle 
= b_n^2,
\qquad
\mbox{for} \quad n=1,2,3,\dots,N-1,
\end{equation}
and the long-time limit of the rms end-to-end distance is given by
\begin{equation}
\label{eqrmsend}
\mu_{\infty} 
= \lim_{t \rightarrow \infty} \langle \mathbf{R}^2 \rangle^{1/2} 
= \sqrt{\sum_{n=1}^{N-1} b_n^2}.
\end{equation}
\end{lemma}

\smallskip

\begin{proof}
Equations (\ref{SDEdef}) describe a system of $3N$ linear SDEs. 
However, the SDEs corresponding to different spatial dimensions are 
not coupled. We therefore restrict our investigation to the behaviour 
of the first coordinates of each vector in~(\ref{rmsbond}). Let us 
arrange the differences of the first coordinates of subsequent beads 
into the $(N-1)$-dimensional vector 
$$
\vec{y}(t) = [r_{2,1}(t)-r_{1,1}(t),
r_{3,1}(t)-r_{2,1}(t),r_{4,1}(t)-r_{3,1}(t),
\dots,r_{N,1}(t)-r_{N-1,1}(t)].
$$
Then SDEs (\ref{SDEdef}) can be rewritten to the system of SDEs
for $\vec{y}(t)$ in the matrix form
$\mbox{d} \vec{y} = A \, \vec{y} \, \mbox{d}t + B \, \mbox{d}\vec{W},$
where $A \in {\mathbb R}^{(N-1) \times (N-1)}$ is a three-diagonal 
matrix given by
\begin{equation*}
A
=
\left(
\begin{matrix}
- \left(\frac{k_1}{\zeta_1} + \frac{k_1}{\zeta_2} \right) & 
\frac{k_2}{\zeta_2} & 0 & 0 & \dots & 0 \\
\frac{k_1}{\zeta_2} & 
- \left( \frac{k_2}{\zeta_2} + \frac{k_2}{\zeta_3} \right) & 
\frac{k_3}{\zeta_3} & 0 & \dots & 0 \\
0 & \frac{k_2}{\zeta_3} & 
- \left( \frac{k_3}{\zeta_3} + \frac{k_3}{\zeta_4} \right) & 
\frac{k_4}{\zeta_4} & \dots & 0 \\
\vdots & \vdots & \vdots & \vdots & \ddots & \vdots \\
0 & 0 & 0 & 0 & \dots & - 
\left( \frac{k_{N-1}}{\zeta_{N-1}} + \frac{k_{N-1}}{\zeta_{N}} \right) \\
\end{matrix} 
\right),
\end{equation*} 
$B \in {\mathbb R}^{(N-1) \times N}$ is a two-diagonal matrix 
given by
\begin{equation*}
B
=
\left(
\begin{matrix}
- \sqrt{\frac{2 k_B T}{\zeta_1}} & \sqrt{\frac{2 k_B T}{\zeta_2}}
& 0 & 0 & \dots & 0 & 0 \\
0 & - \sqrt{\frac{2 k_B T}{\zeta_2}} & \sqrt{\frac{2 k_B T}{\zeta_3}}
& 0 & \dots & 0 & 0 \\
0 & 0 & - \sqrt{\frac{2 k_B T}{\zeta_3}} & \sqrt{\frac{2 k_B T}{\zeta_4}}
& \dots & 0 & 0 \\
\vdots & \vdots & \vdots & \vdots & \ddots & \vdots & \vdots \\
0 & 0 & 0 & 0 & \dots & 
- \sqrt{\frac{2 k_B T}{\zeta_{N-1}}} & \sqrt{\frac{2 k_B T}{\zeta_N}} \\
\end{matrix} 
\right)
\end{equation*} 
and $\mbox{d}\vec{W}$ is $N$-dimensional noise vector
$
\mbox{d}\vec{W} = [\mbox{{\rm d}}W_{1,1},
\mbox{{\rm d}}W_{2,1},\mbox{{\rm d}}W_{3,1},
\dots, \mbox{{\rm d}}W_{N,1}]^T.
$
The stationary covariance matrix, defined by
$$
C = \lim_{t \to \infty} \langle \vec{y} \, \vec{y}^T \rangle,
$$
is the solution of Lyapunov equation~\cite{hotz1987covariance}
$
A C + C A^T + B B^T = 0
$.
It can be easily verified that the unique solution of this
equation is diagonal matrix $C \in {\mathbb R}^{(N-1) \times (N-1)}$ 
with diagonal elements $b_i^2/3,$ $i=1,2,3,\dots,N-1.$ 
Multiplying this result by 3 (the number of coordinates), 
we obtain~(\ref{rmsbond}). The end-to-end distance can
be rewritten as
$
\mathbf{R} = \mathbf{r}_N - \mathbf{r}_1
= \sum_{n=2}^N 
(\mathbf{r}_n - \mathbf{r}_{n-1}).
$
Substituting into~(\ref{eqrmsend}), using~(\ref{rmsbond}) 
and the fact that the stationary covariance matrix $C$ is 
diagonal, we obtain~(\ref{eqrmsend}).
\end{proof}

\subsection{Optimal model refinement in time and space}
\label{sec:invariance}

Lemmas~\ref{lemdg} and~\ref{lemrms} describe theoretical results which have
been derived under slightly different assumptions. Lemma~\ref{lemdg} is
formulated as a property of Algorithm~\ref{algoneiter}, but 
the same result, equation (\ref{eq:SDC}), also holds when we
calculate the self-diffusion coefficient of the SDE formulation 
of the multi-resolution bead-spring model (\ref{SDEdef}).
Algorithm~\ref{algoneiter} is designed in such a way that all
force terms corresponding to springs cancel when the evolution
equation for $\mathbf{r}_G^{i \Delta t}$ is derived
(see equation (\ref{rgequation})). In particular, Lemma~\ref{lemdg} 
holds for any choices of the lengths of timesteps associated 
with different springs.

On the other hand, Lemma~\ref{lemrms} describes the property of 
the SDE formulation (\ref{SDEdef}). If we use a discretized version 
of (\ref{SDEdef}), then we introduce a discretization error. This
error can be made smaller by choosing smaller timesteps. In this section,
we show that the smallest timesteps are only required in the regions
with the highest spatial resolution. We define a family of 
optimal multi-resolution (OMR) models designed to have macroscopic 
properties invariant to resolution regime.

\smallskip

\begin{definition}
\label{defOMR}
Let us consider a bead-spring polymer consisting of $N$ beads of radius
$\sigma$ connected by $(N-1)$ springs with the Kuhn length $b$ and simulated
by $(\ref{EMdiscR})$ with timestep $\Delta t$. Let us divide the polymer
into $R>1$ regions containing $N_j$, $j=1,2,\dots,R$, consecutive
springs, i.e.
\begin{equation}
\sum_{j=1}^{R} N_j = N-1.
\label{Nm1assum}
\end{equation}
The first region contains springs indexed by $\{1,2,\dots,N_1\}$
and the $j$-th region, $j=2,3,\dots,R$, contains springs indexed by 
$\{1+\sum_{k=1}^{j-1} N_k, 2+\sum_{k=1}^{j-1} N_k, \dots, \sum_{k=1}^j N_k\}$.
Let us associate with each region an integer resolution $s_j$, where 
$s_j | s_{j-1}$ or $s_{j-1} | s_{j}$, for $j=2,3,\dots,R$, and
$s_j^2 | N_j$ for $j=1,2,3,\dots,R$, with 
at least one region in resolution $1$ which is the region with the 
finest detail. Larger values of $s_{j}$ represent coarser 
representations of the filament. We define the OMR model 
as the multi-resolution bead-spring model which consists
of $R$ regions of consecutive beads and springs. In the $j$-th region, 
we have $\widetilde{N}_j$ springs with Kuhn length $\widetilde{b}_j$
and associated time steps $\widetilde{\Delta t}_j$ given by
\begin{equation}
\widetilde{N}_j = \frac{N_j}{s_j^2}, \qquad
\widetilde{b}_j = s_j b, \qquad 
\widetilde{\Delta t}_j = s_j^4 \Delta t, \qquad
\widetilde{\sigma}_j = s_j^2 \sigma,
\label{sjscalings} 
\end{equation}
where $\widetilde{\sigma}_j$ is the radius of beads which are 
connected to two springs which have the same Kuhn length $\widetilde{b}_j$.
We assume that the bead radius of beads on region boundaries sharing 
springs with Kuhn lengths $\widetilde{b}_{j-1}$ and $\widetilde{b}_j$ 
is $(\widetilde{\sigma}_{j-1}+\widetilde{\sigma_{j}})/2$,
for $j=2,3,\dots,R.$ Moreover, we assume that the bead radius of 
the first and last bead of the polymer chain are 
equal to $(\sigma + \widetilde{\sigma}_1)/2$
and $(\sigma + \widetilde{\sigma}_R)/2$, respectively.
\end{definition}

\medskip

\noindent
Substituting scalings (\ref{sjscalings}) into (\ref{eq:COM}),
and using (\ref{Nm1assum}), we obtain that the OMR model 
satisfies
$$
\Omega = 
\sigma
+
\sum_{j=1}^R 
\widetilde{\sigma}_j
\widetilde{N}_j
=
\sigma N,
$$
Substituting into (\ref{eq:SDC}), we deduce that the OMR model has the 
same self-diffusion constant as the original detailed model
(given by (\ref{eq:rouseDiff})). Considering the limit $\Delta t \to 0$,
we can use Lemma~\ref{lemrms} and scalings (\ref{sjscalings})
to derive the expected rms end-to-end distance for the filament:
$$
\mu_\infty
= 
\sqrt{\sum_{j=1}^{R} \widetilde{N}_j \widetilde{b}^2_j}
=
\sqrt{(N-1) b^2},
$$
which is again independent of the choice of resolutions
$s_j$, $j=1,2,\dots,R$. As the Kuhn length and bead radius vary 
across resolutions, it is important to consider the numerical stability 
of the model~\cite{Yoshihiro1996stability}. We choose timesteps to 
be sufficiently small so that solutions do not grow exponentially 
large. In discretized equations of Algorithm~\ref{algoneiter},
drift terms appear in the form 
$(k_n/\zeta_n)(\mathbf{r}_{n+1} - \mathbf{r}_{n}) \Delta t_n$,
which is proportional in the $j$-th region of the OMR model to
\begin{equation}
\frac{1}{\widetilde{b}^2_j \widetilde{\sigma}_j}
\big(\mathbf{r}_{n+1} - \mathbf{r}_{n} \big) 
\widetilde{\Delta t}_j.
\label{driftscal}
\end{equation}
Using scalings (\ref{sjscalings}) and assuming that 
$(\mathbf{r}_{n+1} - \mathbf{r}_{n})$ is of the same order
as the Kuhn length $\widetilde{b}_j$, we obtain
that the size of (\ref{driftscal}) scales with $s_j$.
Assuming that $\Delta t$ is chosen in the original fine
scale model so that
$(k/\zeta)(\mathbf{r}_{n+1} - \mathbf{r}_{n}) \Delta t$
is small compared to the Kuhn length $b$, then the drift term
of the OMR model, given by (\ref{driftscal}) is also small
compared to $\widetilde{b}_j,$ the characteristic lengthscale 
of the OMR model in the $j$-th region, $j=1,2,\dots,R.$

Next, we compare the number of calculations made by the original 
detailed single-scale Rouse model with the OMR model. The $j$-th
region has $\widetilde{N}_j$ springs simulated with timestep
$\widetilde{\Delta t}_j$. Using scalings (\ref{sjscalings}),
we obtain that we use $s_j^6$-times fewer calculations in the $j$-th
region by advancing fewer beads over larger timesteps. Assuming
that the computational intensity of the simulation of the
detailed model in each region is proportional to the size of 
the region, $N_j/(N-1)$, we can quantify the fraction of 
computational time which is spent by the OMR model (as compared to 
the detailed model) by
\begin{equation}
\sum_{j=1}^R \frac{N_j}{N-1} \frac{1}{s_j^6}.
\label{fraccost}
\end{equation}
For example, if we coarse-grained the detailed model everywhere
using the integer resolution $s_1 = 2$, then (\ref{Nm1assum}) 
and (\ref{fraccost}) implies that we speed up our simulations 
by the factor of 64.

\subsection{Simulation results}
\label{sec:demonConsist}

In this section we show that simulations of the OMR method match 
the original single-scale Rouse model. We also compare this to 
analytic results predicted from equations~(\ref{eq:SDC}) 
and~(\ref{eq:rmse2e}) for the rms end-to-end distance and the 
self diffusion constant of a filament in an equilibrium state. 
For the detailed model, we choose the parameters:
\begin{equation}
\label{paramvalues}
\Delta t = 0.8 \, \mu\rm{s}, \qquad
b = 60 \, \rm{nm}, \qquad
\mbox{$\sigma$} =  1.2 \, \rm{nm}, \qquad
\end{equation}
where the Kuhn length is chosen to be longer than the persistence 
length of DNA~\cite{hagerman1988flexibility}, and the other parameters 
are chosen arbitrarily. For the remainder of this paper, we shall 
use $k_B = \num{1.4e-23} \rm{J \, K^{-1}}$, $T = 300\,\rm{K}$ 
and $\eta = 1\,\rm{cP},$ the viscosity of water.

\subsubsection{Comparison at equilibrium}
\label{sec:eqbrmComparis}

We compare two resolution regimes for the same system, with 
the single scale model considering the full system in high 
resolution and a multiscale model considering the middle $10\%$ 
of the filament in high resolution and the remainder in low 
resolution. The corresponding parameters of the OMR model are 
given in Table~\ref{tab:consistencyParams}. The OMR model
contains 69 beads connected by 68 springs, while the original
detailed model is given by 501 beads connected by
500 springs.

\begin{table}
\centering
\begin{tabular}{|c|c|c|c|}
\hline
Region & $j=1$ & $j=2$ & $j=3$ \\
\hline
$N_j$ & 225 & 50 & 225 \\
$s_j$ & 5 & 1 & 5 \\
$\widetilde{N}_j$ & 9 & 50 & 9 \\ 
$\widetilde{b}_j$ & $\num{300} \, \rm{nm}$ 
& $\num{60} \, \rm{nm}$ & $\num{300} \, \rm{nm}$ \\
$\widetilde{\Delta t}_j$ & $\num{500}\, \rm{\mbox{$\mu$} s}$ 
& $\num{0.8} \, \mu\rm{s}$ & 
$\num{500} \, \mu\rm{s}$ \\
$\widetilde{\sigma}_j$ & $\num{30}\, \rm{nm}$ 
& $\num{1.2} \, \rm{nm}$ & $\num{30} \, \rm{nm}$ \\
\hline
\end{tabular}
\caption{{\it Parameters of the OMR model system used 
to demonstrate consistency.}
\label{tab:consistencyParams}}
\end{table}

We generate the initial configuration of the polymer filament using 
a multiscale generalisation of the Freely Jointed Chain (FJC)
model~\cite{kuhn1934fjc,guth1934innermolekularen}.
The chain is generated iteratively, with the $(n+1)^{\text{th}}$ bead
in the chain placed uniformly at random on the surface of the 
sphere with radius $b_n$ centred on the $n$-th bead, so that a chain 
is produced with unconstrained random angles~\cite{Andrews2014methods}.
We run the model for one second and estimate both the rms end-to-end 
distance and the self diffusion constant of a filament in an 
equilibrium state. The results given in Table~\ref{tab:consistency}. 
From the results we can see that the OMR model accurately maintains 
macroscopic properties at a fraction of number of calculations of the 
detailed model, where $99.94\%$ of the calculations used are used to 
update beads in the fine resolution region.

The length of the simulated DNA was chosen to be short enough for 
easy computation, but long enough to be comparable to real simulations 
of the DNA behaviour~\cite{robinson2006}. Those simulations showed 
the diffusion of DNA to be a thousand times larger than our simulations, 
this is due in part to neglecting hydrodynamic forces, which gives 
a different form of the expected self diffusion 
constant~\cite{Doi1986theory} compared to~(\ref{eq:SDC}).

\begin{table}
\begin{center}\vspace{1cm}
\begin{tabular}{|c|c|c|c|}
\hline
Model & 
\hbox{\hsize=2.89cm\vbox{\rule{0pt}{4mm}
Self-diffusion \break \rule{0pt}{4mm}
constant [$\mu\rm{m}^2\rm{s}^{-1}$]}} & 
\hbox{\hsize=2.89cm\vbox{\rule{0pt}{4mm}
RMS end-to-end \break \rule{0pt}{4mm}
distance [$\mu\rm{m}$]}} 
& \hbox{\hsize=2.89cm\vbox{\rule{0pt}{4mm}
Number of \break \rule{0pt}{4mm}
calculations}} \\
\hline
Analytic results & $3.66 \times 10^{-4}$ &
 $1.34$ & N/A \\
Detailed model & $3.55 \times 10^{-4}$ &
 $1.36$ & $100\%$ \\
Multiscale & $3.67 \times 10^{-4}$ &
 $1.34$ & $10.006\%$ \\
\hline
\end{tabular}%
\caption{{\it Comparison of results between the original Rouse model, 
the multiscale Rouse model and analytic results~}\cite{Doi1986theory}. 
{\it The number of calculations is given by}~(\ref{fraccost}).
{\it Estimated from one second-long simulation for each system. 
Analytic results given by equations}~(\ref{eq:SDC}){\it and}~(\ref{eq:rmse2e}).
\label{tab:consistency}}
\end{center}
\end{table}

\subsubsection{Compacting the filament}

We have shown that the dynamics of some multiscale systems 
match the analytic results at equilibrium predicted by
equations (\ref{eq:SDC}) and~(\ref{eq:rmse2e}). It is also 
important to show that non-equilibrium dynamics of the detailed system 
can be replicated by the multiscale systems. We consider a filament 
shrunk so that all beads begin at the origin and compare 
how the models expand towards equilibrium. We compare four systems
(C1, C2, C3 and C4) given in Table~\ref{tab:compactingParams} 
which, by construction, have the same rms end-to-end distance, 
but which vary in their level of detail and the number 
of resolutions considered.

\begin{table}
\centering
\begin{tabular}{|c?c?c?c|c?c|c|c|}
\hline
Scheme & C1 & C2 & \multicolumn{2}{c?}{C3} & \multicolumn{3}{c|}{C4} \\
\hline
Region & $j=1$ & $j=1$ & $j=1$ & $j=2$ & $j=1$ & $j=2$ & $j=3$ \\
\hline
$N_j$ & 250 & 250 & 125 & 125 & 100 & 50 & 100 \\
$s_j$ & 5 & 1 & 1 & 5 & 1 & 5 & 1 \\
$\widetilde{N}_j$ & 10 & 250 & 5 & 125 & 4 & 50 & 4 \\ 
$\widetilde{b}_j$ & $\num{300} \, \rm{nm}$ 
& $\num{60} \, \rm{nm}$ 
& $\num{300} \, \rm{nm}$ & 
$\num{60} \, \rm{nm}$ & $\num{300} \, \rm{nm}$ & 
$\num{60} \, \rm{nm}$ & $\num{300} \, \rm{nm}$ \\
$\widetilde{\Delta t}_j$ & $\num{500} \, \rm{\mbox{$\mu$}s}$ 
& $\num{0.8} \, \mu\rm{s}$ 
& $\num{500} \, \rm{\mbox{$\mu$}s}$ 
& $\num{0.8} \, \mu\rm{s}$ 
& $\num{500} \, \rm{\mbox{$\mu$}s}$ 
& $\num{0.8} \, \mu\rm{s}$ 
& $\num{500} \, \rm{\mbox{$\mu$}s}$ \\
$\widetilde{\sigma}_j$ & $\num{30} \, \rm{nm}$ 
& $\num{1.2} \, \rm{nm}$ 
& $\num{30} \, \rm{nm}$ & 
$\num{1.2} \, \rm{nm}$ & 
$\num{30} \, \rm{nm}$ & 
$\num{1.2} \, \rm{nm}$ 
& $\num{30} \, \rm{nm}$ \\
\hline
\end{tabular}
\caption{{\it The OMR systems used for testing the 
multiscale formulation outside of equilibrium in a 
compacted state.}
\label{tab:compactingParams}}
\end{table}

\begin{figure}
\centering
\includegraphics[width=10cm]{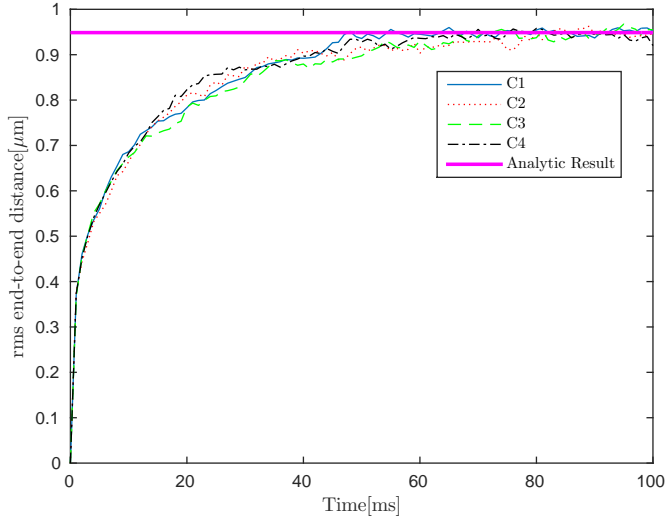}
\caption{{\it The rms end-to-end distance for a compacted filament 
to re-extending towards the analytic (equilibrium) rms end-to-end 
distance in magenta for systems} \rm{C1}--\rm{C4} {\it given in 
Table}~{\rm \ref{tab:compactingParams}}.
\label{fig:compactRes}}
\end{figure}

Model C2 in Table~\ref{tab:compactingParams} is the most detailed
model which uses the same parameters as in (\ref{paramvalues}). 
We compare the time evolution of rms end-to-end distance of
the considered polymer chain models. 
The results (averaged over $10^3$ realizations) are shown  
in Figure~\ref{fig:compactRes}.

\section{DNA binding model}

\label{section4}

To show the usefulness of the OMR model, we have devised an illustrative 
model for the relationship between a DNA binding protein and a segment 
of DNA. There is a large variety of different binding proteins, for 
example transcription factors or polymerases. In general, binding 
protein dynamics are not fully understood and vary greatly depending 
on the function of the protein~\cite{halford2004site}. In Brownian
dynamics models, a binding protein binds to DNA with a given
probability if it comes within a certain distance of the binding
site on the DNA filament~\cite{erban2009stochastic}. We will use
a simple version of this approach and assume that a binding protein 
binds to DNA if it comes within a `binding radius' of a bead in the 
bead-spring polymer chain. 

We apply the multiscale aspect of the model by adjusting the resolution 
regime of the filament dynamically, so that as the binding protein moves 
close to the filament, nearby filament sections increase in resolution. 
To increase the resolution on-the-fly, we develop a Markov Chain Monte 
Carlo (MCMC) scheme. The model is presented as a two-resolution model 
for simplicity, but it can easily be extended to more resolution levels.
We denote by $s$ the difference in resolution between high and low 
regions. Using the notation introduced in Definition~\ref{defOMR}, 
we consider filaments which include regions with higher resolution 
(where $s_j=1$ in Definition~\ref{defOMR}) and regions with lower 
resolution (where $s_j=s$).

\subsection{Resolution increase with MCMC}
\label{sec:ResnIncrease}

We first present a framework for increasing the resolution 
between adjacent beads. If we wish to increase the resolution in a 
region of low resolution between beads $\mathbf{r}_A$ and $\mathbf{r}_B$ 
which lie a distance $d$ apart, then by Definition~\ref{defOMR} we 
introduce $s^2-1$ new springs with Kuhn length $b$.
We place the new beads between $\mathbf{r}_A$ and $\mathbf{r}_B$ to 
lie a distance of $d'$ apart, which is in general not equal to $b$,
because $d$ is in general not equal to $sb$. 
We select $d'$ so that the rms end-to-end distance for the new 
chain formed is to equal $d$ at the time of its creation. 
Using equation~(\ref{eq:rmse2e}), we require
$d' = d/s$ to be the distance between each bead.
It should be noted that we are generating beads by the FJC model, 
introduced in Section~\ref{sec:eqbrmComparis} to be such that the 
distance between each new bead is the same. This is for simplicity as 
when we apply the dynamics there is a fast transition for these bond 
lengths to revert to the equilibrium distribution given by the 
Gaussian chain model~\cite{Doi1986theory}.

In order to apply MCMC, we seek the probability function $\psi$ of 
a chain $F = \{\mathbf{r}_i \, | \, i=1,2,\ldots,M \}$ of $M=s^2-1$ 
beads a distance $d'$ apart between $\mathbf{r}_A$ and $\mathbf{r}_B$. 
We first consider the probability function for the first $M-1$ beads 
$\{\mathbf{r}_{i<M}\} = \{\mathbf{r}_i \, | \, i=1,2,\ldots,M-1 \}$, 
which is proportional to the circumference of the circle of points 
on which the last bead $\mathbf{r}_M$ can be placed, which we call 
the \textit{circle of allowable points} $\Gamma$. This is given 
as the set
\begin{equation}
\nonumber
\Gamma = \{ \mathbf{r} : 
|\mathbf{r}_{M-1} - \mathbf{r}| = d' 
\ \text{and} \  
|\mathbf{r}_B - \mathbf{r}| = d' \},
\end{equation}
or the set of all points a distance $d'$ from both $\mathbf{r}_{M-1}$ 
and $\mathbf{r}_B$. In some cases $\Gamma = \emptyset$ when 
$|\mathbf{r}_{M-1}-\mathbf{r}_B|>2d'$. Note that 
if $\psi(\{\mathbf{r}_{i<M}\})$ is proportional to the circumference 
of circle $\Gamma$ it is also proportional to its radius.

Provided $\Gamma \neq \emptyset$, to find the radius of $\Gamma$ we 
consider the triangle with points $\mathbf{r}_{M-1}$, $\mathbf{u}$ which 
is an arbitrary point on $\Gamma$, and $\mathbf{v}$ which is the midpoint 
between $\mathbf{r}_{M-1}$ and $\mathbf{r}_B$. By construction there 
is a right angle at $\mathbf{v}$. The radius of $\Gamma$ is given by the 
distance between $\mathbf{u}$ and $\mathbf{v}$, which is
$l = \sqrt{d^{\prime2}-(\lambda/2)^{2}},$
where $\lambda$ is the distance between $\mathbf{r}_{M-1}$ 
and $\mathbf{r}_B$. We therefore find the probability density 
function for the chain:
\begin{equation}
\nonumber
\psi(F) = 
\begin{cases}
K \displaystyle
\sqrt{d^{\prime2}-(\lambda/2)^2} 
\prod_{i=1}^{M-1} \delta\Big(|\mathbf{r}_i - \mathbf{r}_{i-1}| - d'\Big), 
& \mbox{if } \Gamma \neq \emptyset; 
\\
0, & \mbox{if } \Gamma = \emptyset;
\end{cases}
\end{equation}
with normalisation constant $K$, $\mathbf{r}_0 = \mathbf{r}_A$ and 
$\delta$ the Dirac delta function. To select from the probability 
distribution $\psi$, we apply the Metropolis-Hastings 
algorithm~\cite{chib1995metropolishastings} as outlined in 
Algorithm~\ref{alg:metHast}.
\begin{algorithm}[t]
Generate a proposal chain 
$\hat{F} = \{\mathbf{\hat{r}}_i \, | \, i=1,2,\ldots,M-1 \}$ with 
each spring length $d'$ starting at $\mathbf{r}_A$ from the 
FJC model. \\
\eIf{$|\mathbf{\hat{r}}_{M-1}-\mathbf{r}_{B}|<2d'$}{
Generate $\nu$ uniformly at random in $[0,1]$. \\
Let $\alpha = \min(\psi(\hat{F})/\psi(F^i),1)$. \\
\eIf{$\nu<\alpha$}{
Pick $\mathbf{\hat{r}}_{M}$ uniformly at random on the circle 
of points distance $d'$ from both $\mathbf{\hat{r}}_{M-1}$ 
and $\mathbf{r}_{B}$. \\
Select $F^{i+1}=\hat{F}$. \\}{
Select $F^{i+1}=F^i$. \\
}}
{Repeat and generate a new proposal chain.}
\caption{{\it The \hbox{\strut}algorithm for picking the next 
chain ${F}^{i+1}$ in the Metropolis-Hastings algorithm given 
the previous chain $F^i$.}
\label{alg:metHast}}
\end{algorithm}%
Given the first $M-1$ beads, and provided $\Gamma \neq \emptyset$, 
the density function for the final bead is given by
$$
\frac{1}{\sqrt{d^{\prime2} - (\lambda/2)^2}} 
\, \delta(|\mathbf{r}_M - \mathbf{r}_{M-1}| - d') 
\, \delta(|\mathbf{r}_B - \mathbf{r}_M| - d').
$$
Algorithm~\ref{alg:metHast} is a Metropolis-Hastings algorithm with 
a candidate-generating distribution which uses rejection 
sampling~\cite{neal2003rejectionsampling} to keep selecting 
chains $\hat{F} = \{\mathbf{\hat{r}}_i \, | \, i=1,2,\ldots,M-1 \}$ 
with distance $d'$ between each bead until we find a chain 
where the second last bead $\mathbf{\hat{r}}_{M-1}$ is within 
a distance $2 d'$ from the endpoint $\mathbf{r}_{B}$, which is 
then used as the candidate. Based on $10^4$ simulations we see 
approximately a $77.5\%$ acceptance of the candidate for $M \geq 3$ 
beads generated between beads a distance $b$ apart, invariant to the 
number of beads for $M \geq 3$. To ensure the Markov chain is 
sufficiently well mixed and that we have `lost memory' of the 
initial chain, we choose the $10^{\text{th}}$ filament in the 
sequence generated from Algorithm~\ref{alg:metHast}; in $10^3$ simulations 
with a resolution difference of $3$, we see only one case where the 
initial distribution is the same as the $10^{\text{th}}$.

\subsection{Model description}

On top of our multiscale Rouse model, we introduce a binding 
protein, modelled as a diffusing particle with diffusion constant 
$D_p$. The position of the binding protein at time $t$ is
denoted $\mathbf{p}(t)$. We assume that it evolves according to 
the discretized Brownian motion
\begin{equation}
\label{eq:particleDiffuse}
\mathbf{p}(t + \Delta t) 
= \mathbf{p}(t) + \sqrt{2 D_p \Delta t} \ \bm{\xi},
\end{equation}
where $\bm{\xi} \sim [\mathcal{N}(0,1),\mathcal{N}(0,1),\mathcal{N}(0,1)]$
and $\Delta t$ is the time step of the highest resolution. 
Associated with the binding protein is a binding radius $d_b$, such 
that if any bead along the filament comes within a distance $d_b$ from 
the binding protein then it `binds' to the filament and the simulation 
finishes. We also consider the protein to have drifted to infinity and 
fail to bind if it gets a distance $d_{\infty}$ from all beads which 
exist in the lowest resolution of the OMR model.

\begin{figure}
\includegraphics[width=0.32\linewidth]{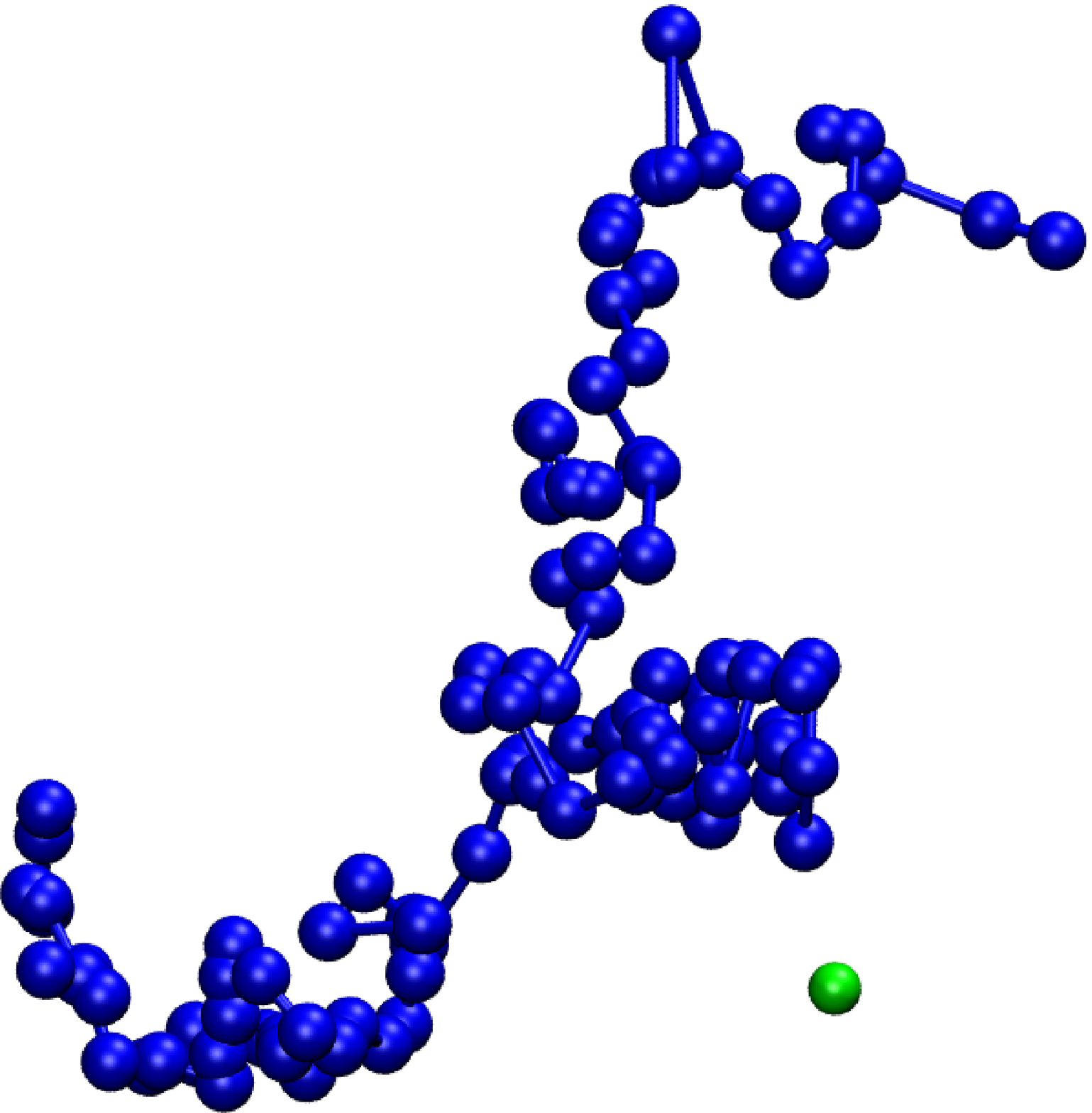}
$\;$
\includegraphics[width=0.32\linewidth]{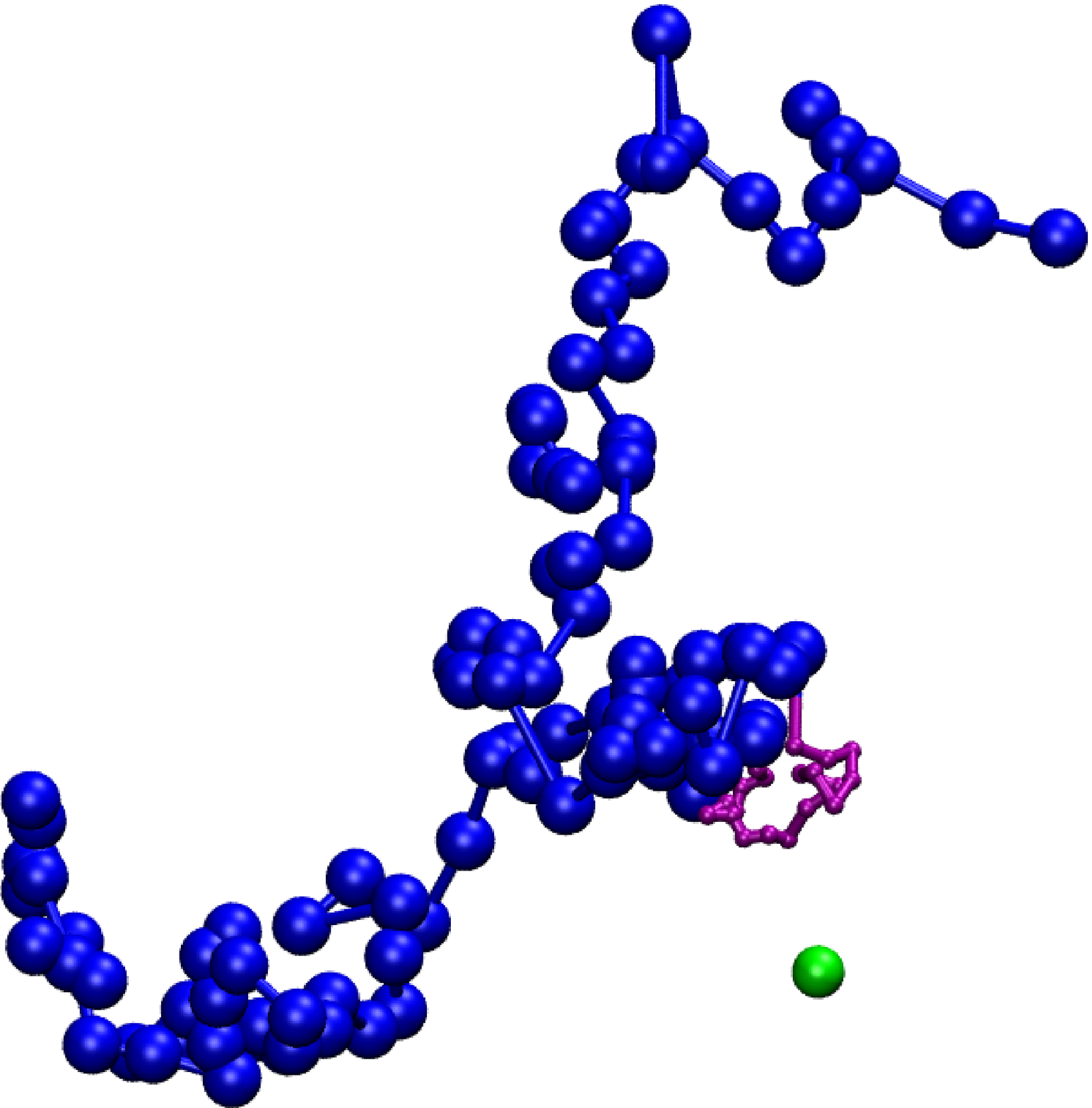}
$\;$
\includegraphics[width=0.32\linewidth]{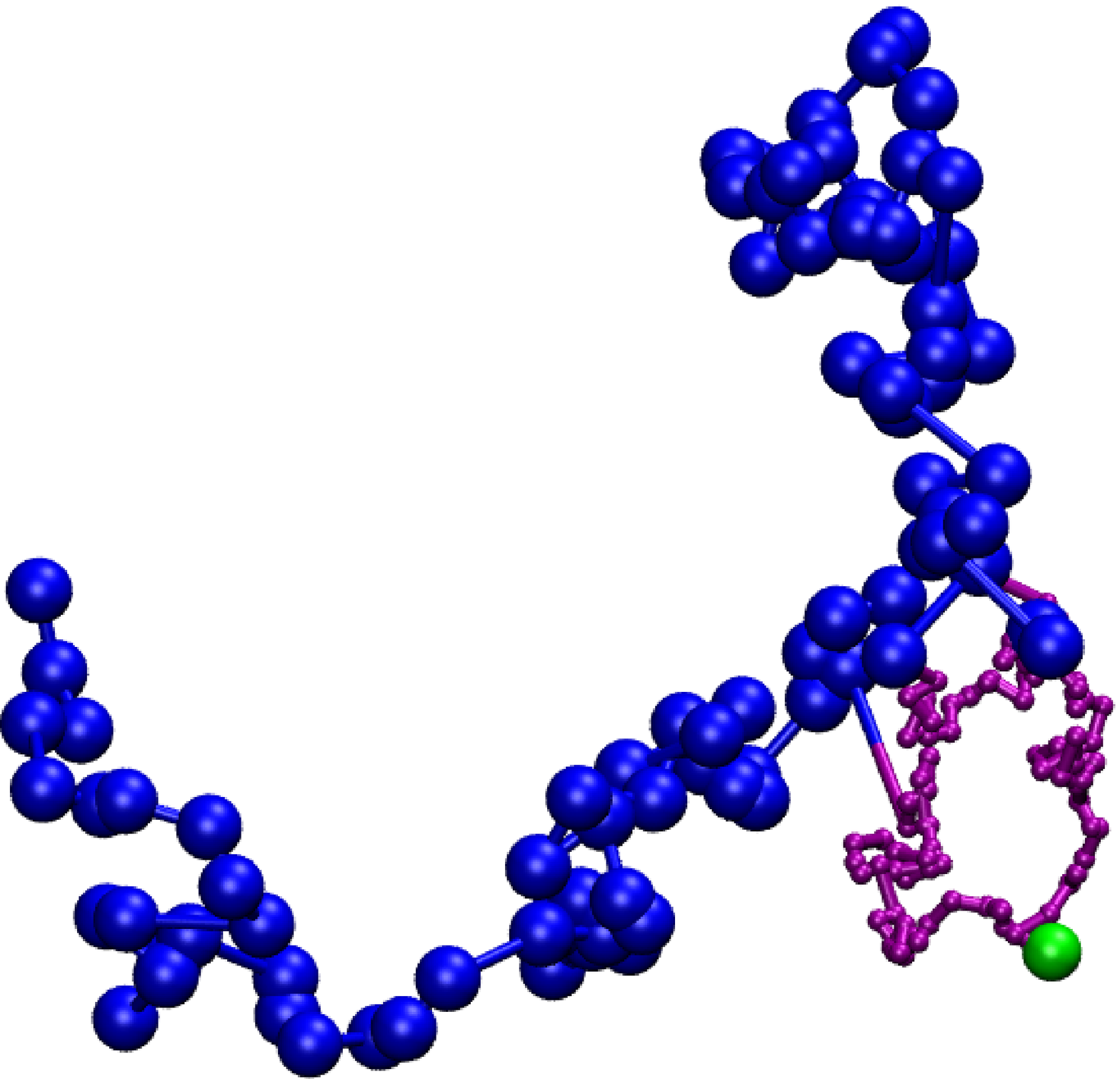}  
\vskip -4.2cm
(a) \hskip 3.8cm (b) \hskip 3.8cm (c)
\vskip 3.8 cm
\caption{{\it Snapshots of the DNA binding model, taken from an
illustrative OMR simulation. 
In} {\rm (a)} {\it the filament is represented by blue in its lowest 
resolution with the binding protein in green. This is a snapshot from 
just before the protein comes within resolution increase radius
$r_I$. In} {\rm (b)} {\it the protein has moved closer than this 
radius, so we increase the resolution in this region. The higher 
resolution region is shown in purple. We continue the simulation 
until either the protein drifts away from the filament or binds to it, 
as we see in} {\rm (c)}.
\label{fig:resIncrease}}
\end{figure}

The binding protein also has its resolution increase radius $r_I$
(where $r_I > d_b$), so 
that if the protein is within a distance $r_I$ from a bead with an 
adjacent spring in low resolution then we change the spring to be 
high resolution and introduce new beads as described in 
Section~\ref{sec:ResnIncrease}, and reformulate bead radius, timestep
and Kuhn length according to the OMR model in 
Definition~\ref{defOMR}. See Figure~\ref{fig:resIncrease} for 
an example of the resolution increasing as the filament comes 
within range of the protein. 

As well as zooming in to specific areas of the filament, we include 
a provision that if the protein moves away far enough from the 
zoomed in filament section that it will not be interacting with 
the filament, then we zoom out. This is implemented as a 
\textit{zoom out factor} $Z$, so that if the protein moves $Z \, r_I$ 
away from both beads on the boundary of a region in high resolution, 
then we `zoom out' by changing the resolution of the region to 
the lower resolution and removing all beads inside.

To initialize the simulation, we generate the filament using the FJC 
model, with $R$ regions each containing one spring and the beads on 
the region boundaries denoted $\mathbf{q}_j(0)$, $j=1,2,\dots,R+1.$
Since all regions are in low resolution at time $t=0$,
scalings (\ref{sjscalings}) are given by
\begin{equation}
\label{coarseregionparam}
\widetilde{N}_j = 1, \qquad
\widetilde{b}_j = s \, b, \qquad 
\widetilde{\Delta t}_j = s^4 \Delta t, \qquad
\widetilde{\sigma}_j = s^2 \sigma,
\end{equation}
for all regions $j=1,2,\dots,R.$
We use $R=100$ in our illustrative simulations.
We then place the protein uniformly at random on the sphere 
with radius $d_0$ centred at the middle bead of the chain
(i.e. at the $(1+R/2)$-th bead). Once the initial configuration has 
been generated, we compute the time evolution of the filament and 
the protein iteratively using Algorithm~\ref{alg:transcription}, which 
describes one timestep of the method. 
\begin{algorithm}[t]
  Calculate the position of the chain at time 
  $t+\Delta t$ by Algorithm~\ref{algoneiter}.
  Update both $\mathbf{r}_n(t+\Delta t)$ for $n=1,2,\dots,N(t)$ and
  $\mathbf{q}_j(t+\Delta t)$ for $j=1,2,\dots,R+1$. \\
  Calculate the protein position at time $t+\Delta t$ by equation~(\ref{eq:particleDiffuse}). \\
  Set $t=t+\Delta t.$ \\
  \uIf{$\min_{1 \leq n\leq N(t)}(|\mathbf{r}_n-\mathbf{p}|)<d_b$}
  {
    The \rule{0pt}{4mm}protein binds to the filament. {\bf STOP} the simulation. \\
  }
  \uElseIf{$\min_{1 \leq j\leq R+1}(|\mathbf{q}_j-\mathbf{p}|)>d_{\infty}$}
  {
    The \rule{0pt}{4mm}protein drifts to infinity. {\bf STOP} the simulation. \\
  }
  \Else{
  \For{$\big(j = 1,2,\dots,R\big)$}
    {
      \uIf{{ \rm (the $j$-th region is in the low resolution)}}
      {
      \If{$\big( |\mathbf{q}_j - \mathbf{p}| < r_I$ $\;$ {\bf OR} $\;$ 
      $|\mathbf{q}_{j+1} - \mathbf{p}| < r_I \big)$}
      {
	Introduce \rule{0pt}{4mm}new beads according to Algorithm~\ref{alg:metHast}. \\ 
	Adjust bead radius, timestep and Kuhn length in the $j$-th region using
	Definition~\ref{defOMR}, i.e. use $\sigma$, $\Delta t$ and $b$, respectively. \\
      }}
      \Else{
      { \rm (the $j$-th region is in the high resolution)} \\
      \If{$\big(|\mathbf{q}_j - \mathbf{p}| > Z \, r_I$ $\;$
      {\bf AND} $\;$ $|\mathbf{q}_{j+1} - \mathbf{p}| > Z \, r_I \big)$}
      {
	Remove \rule{0pt}{4mm}all beads inside the $j$-th region. \\
	Adjust bead radii $\widetilde{\sigma}_j$,
	timestep $\widetilde{\Delta t}_j$ and Kuhn length
	$\widetilde{b}_j$ in the $j$-th region by (\ref{coarseregionparam}). \\
      }}
    }
  }
\caption{{\it One iteration of the \hbox{\strut}algorithm for the DNA binding model.}
\label{alg:transcription}}
\end{algorithm}%
At each time step, the multi-resolution bead-spring polymer is described as 
in Definition~\ref{defmrbs} by the positions of $N \equiv N(t)$
beads at $\mathbf{r}_n \equiv \mathbf{r}_n(t) 
= [r_{n,1}(t),r_{n,2}(t),r_{n,3}(t)]$, $n = 1,2,\dots,N(t).$
We initially have $N(0) = R+1$ beads at positions
$\mathbf{r}_n(0) = \mathbf{q}_n(0)$, for $n=1,2,\dots,N(0).$
As time progresses, some regions are refined, so the value of
$N(t)$ changes and indices of some beads are relabelled. 
To simplify the presentation of Algorithm~\ref{alg:transcription}, 
we denote by $\mathbf{q}_j \equiv \mathbf{q}_j(t)$, 
$j=1,2,\dots,R+1,$ the positions of boundary beads of each region 
at time $t$, independently of the fact whether the chain was 
refined or not in the corresponding region.

We have confirmed the consistency of the model where resolutions 
are changed dynamically against analytic results.  
It performs well with a rms end-to-end distance of $2.997 \, \mu\rm{m}$ 
and diffusion of $7.49\cdot10^{-5}\,\mu\rm{m^2s^{-1}}$ compared to 
expected results from~(\ref{eq:SDC}) and~(\ref{eqrmsend}) of 
$3\,\mu\rm{m}$ and $7.34\cdot10^{-5}\,\mu\rm{m^2s^{-1}}$, 
respectively, after $10^3$ simulations, with parameters 
given in (\ref{paramvalues}) and Table~\ref{tab:transcriptionParmas}.

\subsection{Results}

\begin{figure}
\centering
\includegraphics[width=\linewidth]{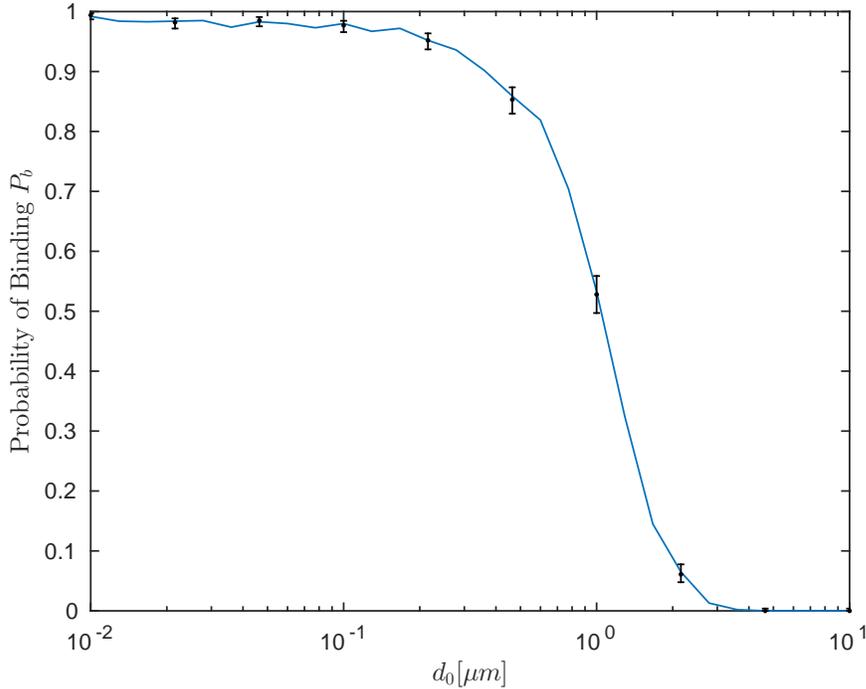}
\caption{{\it A comparison between the probability of binding 
of the protein to the DNA filament, $P_b(d_0)$, depending on 
starting distance, $d_0$, from the filament for the single-scale 
(black points) and OMR (blue line) models. Error bars give a 95\% 
confidence interval based on the Wilson score interval for 
binomial distributions~{\rm \cite{wilson1927probable}}. 
}
\label{fig:transcriptionResults}}
\end{figure}

We compare the results between the OMR model with the classical 
Rouse model, with the entire filament in the highest resolution. 
We run Algorithm~\ref{alg:transcription} with the parameters given 
in (\ref{paramvalues}) and Table~\ref{tab:transcriptionParmas}.
\begin{table}
\begin{center}
\begin{tabular}{ | c | p{3.3cm} | l | p{4cm} | }
  \hline
  \!\!\textbf{Symbol}\!\! & \textbf{Name} & \textbf{Value} 
  & \textbf{Justification} \\ \hline
  $r_I$ & Resolution increase \raggedright{radius} 
  & $0.5 \, \mu\rm{m}$ & Chosen to ensure no instant binding of protein to filament after resolution increase \\ 
  $s$ & Resolution difference & $3$ & Arbitrary \\
  $D_p$ & Protein diffusion constant \raggedright{parameter} 
  & $\num{2e-5}\mu\rm{m^2s^{-1}}$ 
  & Approx. the diffusion constant for the filament \\ \
  $d_b$ & Binding radius & $0.01 \, \mu\rm{m}$ & Arbitrary \\ 
  $d_{\infty}$ & Drift to infinity \raggedright{distance} 
  & $1 \, \mu\rm{m}$ & Arbitrary \\
  $Z$ & Zoom factor & $5$ & Arbitrary \\ 
  $R$ & Number of regions 
      & $100$ & Initial number of beads \\ \hline
\end{tabular}
\caption{{\it Parameter selection for the DNA binding model, 
with the remaining parameters identical to those given 
in}~(\ref{paramvalues}).
\label{tab:transcriptionParmas}}
\end{center}
\end{table}%
At different initial starting distances the model runs until the 
protein is either bound or has escaped the filament. In 
Figure~\ref{fig:transcriptionResults}, we present the probability
of binding of the protein to the filament, $P_b(d_0)$, as a function 
of the initial distance $d_0 \in [d_b,10 \, d_\infty]$ of the 
protein from the middle bead of the filament. We estimate 
$P_b(d_0)$ as a fraction of simulations which end up with the protein
bound to DNA. Each data point in Figure~\ref{fig:transcriptionResults}
represents the value of $P_b(d_0)$ estimated from $10^3$ independent
realizations of the process. If $d_0 < d_b$, then the 
protein is immediately bound to DNA, i.e. $P_b(d_0)=1$ for $d_0 < d_b$. 
If $d_0=d_\infty$, then the probability of binding is nonzero,
because the initial placement, $d_0$, is the distance of the protein 
from the centre of the filament. In particular, the minimum distance 
from protein to filament is less than or equal to the initial 
placement distance, $d_0$, and the simulations (with the possibility
of binding) take place even if $d_0=d_\infty$.
Due to computational constraints of the single-scale model we consider
a selection of initial distances at points $d_0 = 10^{-2+\ell/3} \, \mu$m,
$\ell=0,1,\dots,9$ (black points),
where error bars give a 95\% confidence interval based on the Wilson 
score interval for binomial distributions~{\rm \cite{wilson1927probable}}.
We run simulations for more initial distances, $d_0 = 10^{-2+\ell/9} \, \mu$m,
$\ell=0,1,\dots,27$ (blue line), using the 
computationally efficient OMR model and present our results as the blue
line in Figure~\ref{fig:transcriptionResults}. We see that $P_b(d_0)$ 
is very similar between the single-scale and OMR models. The model also 
succeeds in reducing computational time. For $10^3$ simulations with the 
protein starting $1 \,\mu\rm{m}$ from the middle bead, with parameters 
given in Table~\ref{tab:transcriptionParmas}, the OMR model 
represented a 3.2-times speedup compared to the detailed model, 
with only a 3-times resolution difference. We expect for larger 
resolution differences to see greater improvements in speed.

\section{Discussion}
\label{secdiscussion}

In this paper we have extended basic filament modelling techniques 
to multiple scales by developing OMR methods. We have 
presented an MCMC approach for increasing the resolution 
along a static filament segment, as well as an extension to the 
Rouse model to dynamically model a filament which considers 
multiple scales. The bead radius, as well as the number of beads 
associated with each resolution, is altered to 
maintain consistency with the end-to-end distance and diffusion 
of a filament across multiple scales, as well as the timestep 
to ensure numerical convergence.

We have then illustrated the OMR methodology using a simple
model of protein binding to a DNA filament, in which the
OMR model gave similar results to the single-scale 
model. We have also observed a 3.2-times speed-up in computational 
time on a model which considers only a 3-times increase in 
resolution, which illustrates the use of the OMR approach as 
a method to speed up simulations whilst maintaining the same 
degree of accuracy as the more computationally intensive 
single-scale model. The speed-up in computational time could be 
further increased by replacing Brownian dynamics based on 
time-discretization (\ref{eq:particleDiffuse}) by event-based 
algorithms such as the FPKMC (First passage kinetic Monte Carlo) 
and  GFRD (Green's function reaction dynamics) 
methods~\cite{Opplestrup:2009:FKM,Takahashi:2010:STC}.

When considering the zooming out of the DNA binding model, 
note that it is generally possible to zoom in and out 
repetitively, as long as the dynamics are such that we can 
generate a high resolution structure independent from the 
previous one (i.e., once we zoom out, the microscopic structure 
is completely forgotten). However, particularly in the case of 
chromatin, histone modification and some DNA-binding proteins 
may act as long-term memory at a microscopic scale 
below the scales currently considered. To reflect the effect 
of the memory, some properties of the microscopic structure 
should be maintained even after zooming out. Fractal dimension 
may serve as a candidate of indices~\cite{mirny2011fractal}, which 
can be also estimated in living cells by single-molecule tracking
experiments~\cite{shinkai2016}.

The OMR method could be applied to modern simulations of DNA and 
other biological polymers which use the Rouse model~\cite{hur2000brownian} 
in situations where certain regions of the polymer require higher 
resolutions than other regions. The model considered in this report 
uses Rouse dynamics, which is moderately accurate given its simplicity, 
but as we zoom in further towards a binding site, then we will need to 
start to consider hydrodynamic forces and excluded volume effects 
acting between beads. Models which include hydrodynamic interactions 
such as the Zimm model~\cite{Zimm1959Model} have previously been used 
to look at filament dynamics~\cite{allison1986brownian,Ermak1978Brownian}. 
Therefore it is of interest to have a hybrid model which uses 
the Rouse model in low resolutions and the Zimm model in high resolutions. 
The combination of different dynamical models might give interesting 
results regarding hierarchical structures forming as we move between 
resolutions.

As we go into higher resolutions, strands of DNA can be modelled as 
smooth~\cite{Andrews2014methods}, unlike the FJC model where angles 
between beads are unconstrained. The wormlike chain model of Kratky 
and Porod~\cite{Kratky1949WLC}, implemented via algorithm by Hagermann 
and Zimm~\cite{Hagermann1981StaticWLC}, gives a non-uniform probability 
distribution for the angles between each bead. Allison~\cite{allison1986brownian} 
then implements the Zimm model dynamics on top of the static formulation 
to give bending as well as stretching forces. Another interesting
open multiscale problem is to implement this at higher resolutions, 
with the Rouse model at lower resolutions, in order to design 
a hybrid model.

To introduce even more realism, we would see individual histones and 
consider forces between these as in the model of Rosa and 
Everaers~\cite{rosa2008structure} which includes Lennard-Jones and 
FENE forces between beads. As we approach an atomistic level, it may 
be interesting to consider a molecular dynamics approach to modelling 
the DNA filament. Coarser Brownian dynamics models can be estimated
from molecular dynamics models either analytically~\cite{erban2014molecular}
or numerically~\cite{couplingallatom}, depending on the complexity
of the molecular dynamics model. A variety of structure-based coarse-grained 
models have been used for chromatin (e.g.~\cite{mueller2014model}), also 
with transcription factors~\cite{takada2015review}. Multiscale modelling 
techniques (e.g.~\cite{korolev2016review} with iterative coarse-graining), as 
well as adaptive resolution models (e.g.~\cite{zavadlav2015} for solvent 
molecules), have been developed. We expect these studies will connect 
with polymer-like models at a certain appropriate length and time scale. 
On top of this, models for the target searching process by proteins 
such as transcription factors could be improved (for example, by
incorporating facilitated diffusion under crowded 
environment~\cite{brackley2013}).

The need for developing and analyzing multiscale models of DNA which
use one of the above detailed simulation approaches for small parts 
of the DNA filament is further stimulated by recent experimental results. 
Chromosome conformation capture (3C)-related techniques, particularly at 
a genome-wide level using high-throughput sequencing (Hi-C~\cite{lieberman2009hic}), 
provide the three-dimensional structure of the chromosomes in an averaged manner. 
Moreover, recent imaging techniques have enabled us to observe simultaneously 
the motion and transcription of designated gene loci in living 
cells~\cite{ochiai2015rolex}. Simulated processes could be compared 
with such experimental results. Recent Hi-C experiments also revealed fine 
structures such as loops induced by DNA-binding proteins~\cite{sanborn2015}. To 
develop more realistic models, information about the binding sites for these 
proteins may be utilized when we increase the resolution in our scheme.

\end{document}